\documentclass[aps,pre,reprint,groupedaddress,showkeys,showpacs]{revtex4-1}
\usepackage{graphicx}
\usepackage{epstopdf}
\usepackage{lineno}
\usepackage[dvipsnames]{xcolor}
\usepackage{amssymb}                    

\begin{document}

\title{Sign singularity of the local energy transfer in space plasma turbulence}

\author{Luca Sorriso-Valvo}
\affiliation{Nanotec/CNR, U.O.S. di Cosenza, Ponte P. Bucci, cubo 31C, 87036 Rende, Italy}
\affiliation{Departamento de F\'isica, Escuela Polit\'ecnica Nacional, Quito, Ecuador}
\author{Gaetano De Vita}
\affiliation{Nanotec/CNR, U.O.S. di Cosenza, Ponte P. Bucci, cubo 31C, 87036 Rende, Italy}
\author{Federico Fraternale}
\affiliation{Dipartimento di Scienza Applicata e Tecnologia, Politecnico di Torino, Torino, Italy}
\author{Alexandre Gurchumelia}
\affiliation{M. Nodia Institute of Geophysics, Iv. Javakhishvili Tbilisi State University, Tbilisi, Georgia}
\author{Silvia Perri}
\affiliation{Dipartimento di Fisica, Universit\`a della Calabria, Rende, Italy}
\author{Giuseppina Nigro}
\affiliation{Dipartimento di Fisica, Universit\`a della Calabria, Rende, Italy}
\author{Filomena Catapano}
\affiliation{Serco Italia for ESA-ESRIN, Frascati, Italy}
\author{Alessandro Retin\`o}
\affiliation{LPP-CNRS/Ecole Polytechnique/Sorbonne Universit\'e, Paris, France}
\author{Christopher H. K. Chen}
\affiliation{School of Physics and Astronomy, Queen Mary University of London, UK}
\author{Emiliya Yordanova}
\affiliation{Swedish Institute of Space Physics, Uppsala, Sweden}
\author{Oreste Pezzi}
\affiliation{Gran Sasso Science Institute, Viale F. Crispi 7, 67100 L’Aquila, Italy}
\affiliation{INFN/Laboratori Nazionali del Gran Sasso, Assergi, Italy}
\author{Khatuna Chargazia}
\affiliation{M. Nodia Institute of Geophysics, Iv. Javakhishvili Tbilisi State University, Tbilisi, Georgia}\affiliation{I. Vekua Institute of Applied Mathematics, Iv. Javakhishvili Tbilisi State University, Tbilisi, Georgia}
\author{Oleg Kharshiladze} 
\affiliation{M. Nodia Institute of Geophysics, Iv. Javakhishvili Tbilisi State University, Tbilisi, Georgia}
\author{Diana Kvaratskhelia} 
\affiliation{M. Nodia Institute of Geophysics, Iv. Javakhishvili Tbilisi State University, Tbilisi, Georgia}\affiliation{Sokhumi State University, Tbilisi, Georgia}
\author{Christian L. V\'asconez}
\affiliation{Departamento de F\'isica, Escuela Polit\'ecnica Nacional, Quito, Ecuador}%
\author{Raffaele Marino}
\affiliation{Laboratoire de M\'ecanique des Fluides et d'Acoustique, CNRS, \'Ecole Centrale de Lyon, Universit\'e Claude Bernard Lyon~1, INSA de Lyon, F-69134 \'Ecully, France}
\author{Olivier Le Contel}
\affiliation{LPP-CNRS/Ecole Polytechnique/Sorbonne Universit\'e, Paris, France}
\author{Barbara Giles}
\affiliation{NASA, Goddard Space Flight Center, Greenbelt MD 20771, USA}
\author{Thomas E. Moore}
\affiliation{NASA, Goddard Space Flight Center, Greenbelt MD 20771, USA}
\author{Roy B. Torbert}
\affiliation{Space Science Center, University of New Hampshire, Durham, New Hampshire, USA}
\author{James L. Burch}
\affiliation{Southwest Research Institute, San Antonio, Texas, USA}

\date{\today}

\begin{abstract}
In weakly collisional space plasmas, the turbulent cascade provides most of the energy that is dissipated at small scales by various kinetic processes. Understanding the characteristics of such dissipative mechanisms requires the accurate knowledge of the fluctuations that make energy available for conversion at small scales, as different dissipation processes are triggered by fluctuations of a different nature.
The scaling properties of different energy channels are estimated here using a proxy of the local energy transfer, based on the third-order moment scaling law for magnetohydrodynamic turbulence. In particular, the sign-singularity analysis was used to explore the scaling properties of the alternating positive-negative energy fluxes, thus providing information on the structure and topology of such fluxes for each of the different type of fluctuations. The results show the highly complex geometrical nature of the flux, and that the local contributions associated with energy and cross-helicity nonlinear transfer have similar scaling properties. Consequently, the fractal properties of current and vorticity structures are similar to those of the Alfv\'enic fluctuations.
\end{abstract}



\maketitle

\section{Introduction}

The dynamics of space plasmas is characterized by a broad variety of complex processes that include turbulence, instabilities, and several mechanisms of particle-radiation interaction. Such processes are intrinsically connected across multiple scales. For example, the energy associated with large-scale structures and instabilities is transported towards smaller and smaller scales though a turbulent cascade due to the nonlinear interactions among magnetic and velocity fluctuations, throughout the so-called inertial range that may span one to more than three decades in scales~\cite{frisch,tumarsch,livingreviews}. 
When the energy reaches scales of the order of or smaller than the typical ion and electron scales (e.g. the proton Larmor radius or inertial length),
a different turbulent cascade occurs~\cite{macek,chenboldyrev}. At those scales, weakly collisional plasma kinetic processes arise, such as nonlinear damping of waves, kinetic instabilities, particle collisions, and magnetic reconnection, that convert the energy stored in the field fluctuations into particle energization and acceleration, and plasma heating~\cite{salem2012,chang2015,chen16,pezzi2016,pezzi2017,chen2019}. 

Past theoretical, experimental and numerical attempts to describe these processes have focused mostly on simplified, idealized conditions. However, in recent years there is an increasing interest in their cross-scale, interwoven nature. Multi-spacecraft and high resolution measurements in the solar wind and in the terrestrial magnetosphere~\cite{cluster,mms} have provided evidence of such interconnection~\cite{grecoperri}. The increasing performance of numerical simulations has also allowed processes on several scales to be examined, and therefore to highlight their relationship~\cite{karimabadi,valentini2014,franci2015,servidio2015,valentini2016,cerri2017,groselj2017,franci2018,pezzi2018,perrone2018}. 
Theoretical efforts are also being carried out in order to highlight the specific processes governing the energy exchange between ranges associated to different regimes~\citep{schekochihin,servidio17,sorriso2018a}.
In this framework, the local, fine-scale details of the turbulent energy cascade acquire new importance, as the specific characteristics of the fluctuations carrying energy to the kinetic scales can be associated with different plasma processes~\citep{klein,sorriso2019,chen2019}. 

Recent analysis has revealed that the temperature and energized particles are enhanced in the proximity of current sheets~\cite{osman,servidioPVI,vasconez15,pucci16,valentini17} or of locations of concentration of turbulent energy~\cite{sorriso2018a,sorriso2018b}. 
The local, fine details of the energy transfer process in the turbulent cascade may therefore play a fundamental role in the activation of those plasma kinetic processes that are believed to be responsible for energy conversion, usually (and loosely) referred to as dissipation.
In numerical simulations specific techniques, mostly based on Fourier-space filtering, have been developed to achieve a detailed description of the energy transfer~\cite{camporeale,yang,kuzzay}. However, the limitations arising from the one-dimensional nature of spacecraft sampling require the introduction of approximated quantities. A simple example is provided by the normalized magnitude of the small-scale magnetic field fluctuations, basically locating current sheets and similar magnetic structures. Techniques known as local intermittency measure (LIM)~\cite{lim} and partial variance of increments (PVI)~\cite{pvi} were extensively used in the last decades. 
A more informative proxy, called local energy transfer (LET), is based on the third-order scaling law for turbulent plasmas~\cite{pp98}, and carries information about the nature of the fluctuations transporting the energy to small scales~\cite{sorriso2019}. For example, the use of this proxy allowed the identification of specific ion features, such as beams, where the alignment between small-scale magnetic field and velocity fluctuations was dominating. This suggested nonlinear resonance between Alfv\'enic fluctuations and particles as a possible mechanism for the generation of those beams~\cite{sorriso2019}.

In this article, the topological properties of the energy flow channels are examined. Measurements from the solar wind and from different regions of the terrestrial magnetosphere are studied by means of sign-singularity analysis. The results show the presence of interwoven positive-negative energy flux, allowing estimation of the typical fractal dimension of the structures, and eventually the role of their different components, in the turbulent cascade and, therefore, on feeding small-scale dissipative processes. Section~\ref{methods} describes the proxy used in this work and the cancellation analysis technique. In Section~\ref{data} we describe the data used. Section~\ref{results} provides a description of the results and the comparison between different data sets. Finally, the results are briefly discussed in Section~\ref{conclusions}.

\section{Methods}
\label{methods}

\subsection{A proxy of the local energy transfer in turbulence}

The fluctuations observed in magnetohydrodynamic plasma turbulence have been shown to follow the Politano-Pouquet law~\citep{pp98}. This predicts the linear scaling of the mixed third-order moment of the fields fluctuations on the scale, when homogeneity, scale separation, isotropy, and time-stationarity are met.
Using the Taylor hypothesis~\citep{taylor,perritaylor} $\mathbf{r}=t\langle \mathbf{v}\rangle$ (necessary to transform space ($\mathbf{r}$) and time ($t$) arguments via the bulk speed $\langle \mathbf{v}\rangle$), the Politano-Pouquet law can be written as 
\begin{equation}
Y({\Delta t}) = \langle 
\Delta v_{l}(|\Delta \mathbf{v}|^2+|\Delta \mathbf{b}|^2)-2\Delta b_{l}(\Delta \mathbf{v}\cdot\Delta \mathbf{b}) \rangle = -\frac{4}{3}  \langle \varepsilon \rangle \Delta t \langle v \rangle \; .
\label{yaglom}
\end{equation}
The mixed third-order moment $Y({\Delta t})$ is computed using the increments $\Delta\psi(t,\Delta t)=\psi(t+\Delta t)-\psi(t)$ of a field $\psi$ (either the plasma velocity $\mathbf{v}$ or the  magnetic field $\mathbf{b}=\mathbf{B}/\sqrt{4\pi\rho}$ given in velocity units through the mass density $\rho$) across a temporal scale $\Delta t$, the subscript $l$ indicating the longitudinal component, {\it i.e.} parallel to the bulk speed. The total energy flux given in equation~(\ref{yaglom}) is proportional to the mean energy transfer rate $\langle\varepsilon\rangle$.  
The Politano-Pouquet law describes the scaling of the  small imbalance between positive and negative energy flux in the turbulent cascade, and is associated with the scale-dependent intrinsic asymmetry (skewness) of the turbulent fluctuations~\cite{frisch,pp98}. 
The linear scaling~(\ref{yaglom}) was robustly observed in numerical simulations~\citep{sorriso2002,andres,ferrand}, in the solar wind plasma~\citep{macbride2005,prl,apjl,coburn,supratik}, and in the terrestrial magnetosheath~\citep{hadid,bandy,bandy2}.
In order to attempt a description of the local energy flux from space data time series, the law~(\ref{yaglom}) can be revisited without computing the average, thus giving a time series of the heuristic proxy of the local energy transfer rates (LET) at a given scale ${\Delta t}$, which can be estimated by computing the quantity: 
\begin{equation}
\varepsilon^\pm(t,{\Delta t}) = -\frac{3}{4} \frac{\Delta v_{l}(|\Delta \mathbf{v}|^2+|\Delta \mathbf{b}|^2)-2\Delta b_{l}(\Delta \mathbf{v}\cdot\Delta \mathbf{b})}{{\Delta t} \langle v \rangle} \, .
\label{pseudoenergy}
\end{equation} 
This procedure neglects several contributions to the scaling, which in~(\ref{yaglom}) are suppressed by averaging over a large sample, and therefore provides only a rough approximation of the actual local energy transfer rate~\cite{hellinger,kuzzay}. 
However, because of the intrinsic difficulty in estimating the neglected terms from one-dimensional data, this proxy can be used as a first degree approximation in space plasmas time series. 
The LET was previously used to determine heating regions in the interplanetary plasma~\citep{sorriso2018a} and on kinetic numerical simulations~\citep{sorriso2018b,yang}.

The LET is composed of two additive terms, one associated with the magnetic and kinetic energy advected by the velocity fluctuations, $\varepsilon_e=-3/(4\Delta t \langle v \rangle) [\Delta v_{l}(\Delta \mathbf{v}^2+\Delta \mathbf{b}^2)]$, and the other with the cross-helicity coupled to the longitudinal magnetic fluctuations, $\varepsilon_c=-3/(4\Delta t \langle v \rangle)[-2\Delta b_{l}(\Delta \mathbf{v}\cdot\Delta \mathbf{b})]$~\citep{sorriso2018a}. Such separation has been used to identify regions dominated by current and vorticity structures from regions dominated by coupled, Alfv\'enic fluctuations in the terrestrial magnetospheric boundary layer, revealing the presence of ion beams mostly associated with the small-scale Alfv\'enic fluctuations, and thus indicating a possible mechanism for the transfer of the turbulent energy to the particles~\citep{sorriso2019}. 
Since the LET, as well as its two separated components, are signed quantities, it may be interesting to explore the scaling properties of the mixing of the positive and negative parts of the turbulent cascade. These may be related to the direction of the energy flow, although this interpretation is not supported by theoretical evidence. Moreover, unlike in the averaged Politano-Pouquet law, decoupling the sign dependence on scale and position is not trivial for the local proxy. Therefore, caution should be used in evaluating the physical meaning of the sign. However, it could still be associated to injection or removal of energy from specific locations and scales. 
It has been shown both in MHD~\cite{Debliquy} and in hydrodynamic flows~\cite{Sahoo} that a selective filter of the triads carrying the energy throughout the inertial range, as well as the absence of resonant triads in the anisotropic case in the presence of rotation and stratification~\cite{Marino2013}, may lead to the modulation of an inverse cascade in fully developed three-dimensional turbulence.
It would be thus interesting to investigate the nature of the sign of the local energy dissipation obtained
with the proxy proposed here also by the implementation of shell models~\cite{Verma,Dar}.

Preliminary comparison between the proxy and more comprehensive estimates of the local energy transfer rate, performed using three-dimensional MHD numerical simulations~\cite{kuzzay}, suggests good qualitative agreement (not shown) in terms of location of the larger transfer regions, although there are some discrepancies in the magnitude and fluctuation of the signed transfers that may relate to the approximated and unfiltered nature of the LET. 
For the purposes of this study, the proxy does not necessarily need to fully capture the turbulent energy flux, as it is rather related to the specific features of the plasma and field fluctuations that contribute to the actual energy flux.

The complexity of the energy flow across scales might carry information about the topology of the small-scale structures, and also specifically for their energy or cross-helicity contributions. This information can be useful to determine which dissipative processes are selectively activated by the turbulent cascade. 
In this work, we aim at providing such information, that will be obtained by means of the cancellation analysis, which is briefly described in the following section.

\subsection{Sign singularity and cancellation analysis}

The properties of chaotic flows can be described through the singularity analysis of the field~\citep{ott}. In particular, if a given field changes sign on arbitrarily small scale, its measure is called {\it sign singular}~\citep{ott}. The quantitative description of this singularity is important for the description of the topological properties (e.g. fractal dimension, filling factor...) of sign-defined (smooth) coherent structures, such as the ones emerging in intermittent, turbulent flows.
A standard technique to estimate sign singularity is provided by the cancellation analysis, previously used to describe the scaling properties of MHD, Hall-MHD and Vlasov-Maxwell turbulence in numerical simulations~\citep{sorriso2002,pouquet,martin,vlasov,gingell} and in the current helicity in solar photospheric active regions~\citep{abra,prelude,sorriso2015,themis}. 

Given a scalar field $f({\bf r})$ with zero mean, defined on a $d$-dimensional domain $Q(L)$ of size~$L$, its signed measure can be defined as the normalized field integrated over scale dependent subsets $Q(l) \subset Q(L)$ of size~$l$, $$\mu(l)=\frac{\int_{Q(l)}\;\mathrm{d} {\bf r} \; f ({\bf r})}{\int_{Q(L)}\;\mathrm{d} {\bf r} \; |f({\bf r})|}\, .$$
A coarse-graining of the domain provides an estimate of the sign-singularity of the measure by  means of the scaling exponent $\kappa$ (also called cancellation exponent) of the cancellation function, which is in turn defined as $\chi(l) =\sum_{Q_i(l)} |\mu_i(l)| \sim l^{-\kappa}$, the sum being intended over all disjoint subsets~$Q_i(l)$ fully covering the domain~$Q(L)$. 
In a chaotic field, positive and negative fluctuations cancel each other if the integral is performed over large subsets, resulting in a small signed measure at large scales. However, if the integration subset has the typical size of the smooth structures, cancellations are reduced and the signed measure is relatively larger. 
The scaling law of the cancellation function, as described by the cancellation exponent, can thus provide information on the field cancellations across the scales. Some specific values of the cancellation exponent can help to interpret the results. If the field is smooth, then the cancellation function does not depend on the scale, and $\kappa=0$. If the field is homogeneous with random discontinuities, then cancellations are enhanced and $\kappa=d/2$. Values in between these two examples indicate the coexistence of random fluctuations and smooth structures, whose fractal dimension $D$ is thus given by $\kappa=(d-D)/2$~\citep{sorriso2002}. The fractal dimension $D$ provides information about the space filling and complexity of the structures carrying the energy to small scales, and might be related to the efficiency of the transport mechanism. In this study we will make use of $\kappa$ and $D$ as parameters to describe the topological properties of the different turbulent energy channels, and compare the results for the interplanetary space and the magnetosphere.

\section{Data}
\label{data} 

In order to study the cancellation properties of the  local energy transfer rate proxy LET, and of its two components, we have selected two magnetospheric plasma intervals measured by the Magnetospheric Multiscale  mission (MMS)~\cite{mms}, which provides data at high cadence, and one longer interval of fast solar wind measured by the Wind spacecraft~\cite{wind}.

The first sample, labeled as MMS-KH, was recorded on September 8, 2015 between 10:07:04
and 11:25:34 UTC, while MMS was in the dusk-side
magnetopause, moving across a portion of plasma dominated by the Kelvin-Helmholtz instability (KH) formed at the boundary between the magnetosheath and the magnetosphere. The interval was extensively studied in the past, showing the presence of strong turbulence and intermittency~\cite{stawarz}. 
The MMS spacecraft performed multiple crossings of the KH boundary, resulting in the alternate sampling of plasma from the magnetosheath and from the magnetospheric boundary layer. The boundary crossings between the two regions are clearly highlighted by sharp transitions of the plasma parameters, so that it is easy to separate them. In this work, we have accurately selected 53 short intervals (ranging one to ten minutes) that are purely immersed in the magnetospheric boundary layer (based on plasma temperature and density), rejecting magnetosheath and transition regions. This allows some degree of homogeneity of the sample, necessary for statistical analysis. 

The second magnetospheric interval, named MMS-MS, was selected in the turbulent magnetosheath region under quasi-parallel bow shock geometry on November 30, 2015, between 00:21 and 00:26 UTC. This 5-minute interval is characterized by intense fluctuations in all plasma and field parameters and by the presence of small scale magnetic structures. 
Some of these have been studied in detail, and various kinetic processes, such as local electron acceleration and magnetic reconnection at thin current sheets have been observed~\cite{eriksson,yordanova,voros}. 

For both MMS intervals, the magnetic field data with sampling frequency $1$ kHz used here are a merged product~\cite{fischer} from the burst mode flux gate (FGM)~\cite{russell} and search coil (SCM)~\cite{lecontel} instruments on MMS. The ion moments come from the fast plasma instrument (FPI)~\cite{pollock} at a sampling rate of $150$ ms.

Finally, in order to compare the magnetospheric results with the solar wind, we have also studied the cancellation properties of the proxy LET using one sample of fast solar wind measured by the Wind spacecraft~\cite{wind}, labeled as WIND. The data interval consists of 6 days when Wind was in a fast stream during days 14 to 19 of 2008, and is the same interval as studied in~\citet{wicks2013}. The data from the magnetic field instrument MFI~\citep{lepping1995} and plasma instrument 3DP~\citep{lin1995} at $3$ s cadence were used, with the magnetic field converted to Alfv\'en units using the kinetic normalization described in~\citet{chen2013}. The average conditions during the interval were a solar wind speed of 660 km/s, magnetic field strength B = 4.4 nT, density 2.4 cm$^{-3}$ and proton beta $\beta_p=1.2$, typical for the fast solar wind.
%
%
%

\begin{figure*}[h!]
\includegraphics[height=0.3\linewidth]{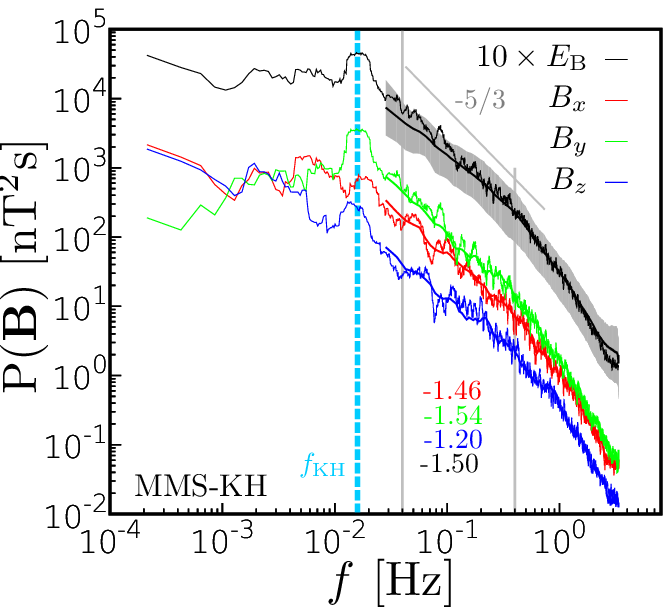}
\includegraphics[height=0.3\linewidth]{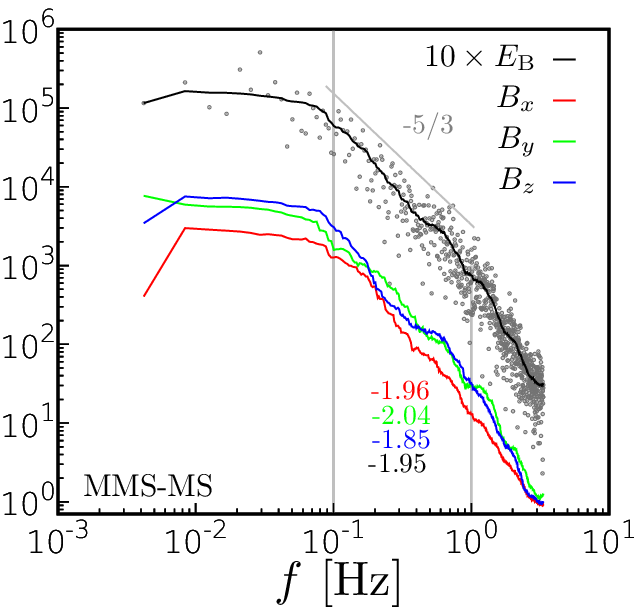}
\includegraphics[height=0.29\linewidth]{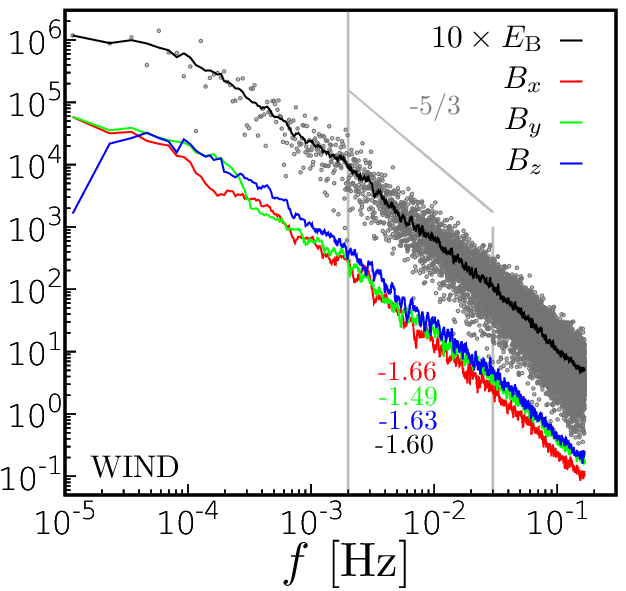}
\caption{Magnetic components spectral power density for the three samples. The trace is shown in black, scales by a factor of ten for clarity. 
For MMS-KH (left panel), both the compressed sensing spectra (thin curves) and the spectra averaged over the sub-intervals (thicker curves) are shown. The standard deviation of all 53 sub-intervals is shown in gray for the trace. The blue-dashed vertical line corresponds to the large-scale KH frequency $f_\mathrm{KH}=0.0146$ Hz. 
For MMS-MS and WIND (central and right panels), the unsmoothed trace is shown with gray points. Fitted power-law exponents in the MHD-inertial frequency range indicated by grey vertical lines are given. Reference power laws with $-5/3$ exponent are also shown.}
\label{fig:spectra}
\end{figure*}

The three intervals are characterized by variable levels of magnetic fluctuations. As shown in Figure~\ref{fig:spectra}, all three intervals present a reasonably well defined power-law spectral scaling range. 
For the WIND and MMS-KH intervals, the spectra are close to the Kolmogorov prediction, with scaling exponents $\sim-5/3$ compatible with the standard values for MHD turbulence. The MMS-KH data shows slightly shallower spectra, but still in the standard range of observation of turbulent space plasmas. Note that the power spectra in these data were obtained using the compressed sensing technique described in~\citet{fraternale}. 
The magnetosheath interval, MMS-MS, has less defined power-law scaling, possibly because of its short duration, and the scaling exponent is $\sim-2$ (see the fitted exponents inside each panel frame), suggesting the presence of uncorrelated structures. This is typical of the highly fluctuating magnetosheath magnetic field, and indicates a relatively less developed turbulence.
In the magnetosheath flanks, Kolmogorov-type power spectra can be observed in the MHD range~\cite{alexandrova,huang}. However, in the region closer to the subsolar point, where MMS orbit lies during this particular event, the plasma is highly compressed and closely confined between the bow shock and the magnetopause. The solar wind turbulence, once modified and shuffled by the bow shock crossing, does not have enough space and time to reach a fully developed state, because of the close proximity of the two large boundaries. This results in the observed steeper spectral exponents. Note that in the present sample the typical ion frequencies are of the order of $1.3$ Hz, which exclude the possibility that the observed scaling range is in the kinetic regime~\cite{macek}.
\begin{figure*}[h!]
\includegraphics[height=0.3\linewidth]{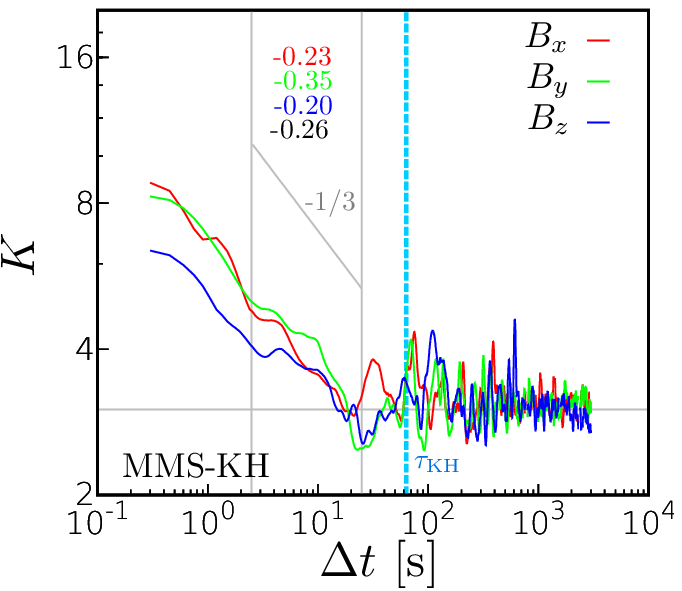}
\includegraphics[height=0.3\linewidth]{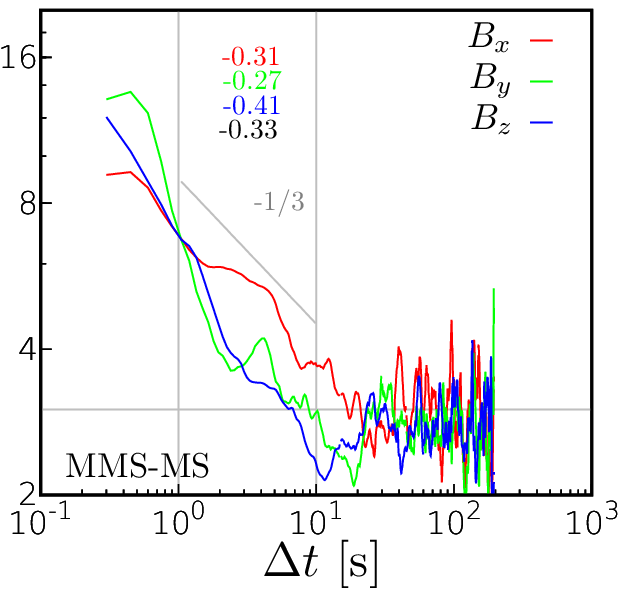}
\includegraphics[height=0.3\linewidth]{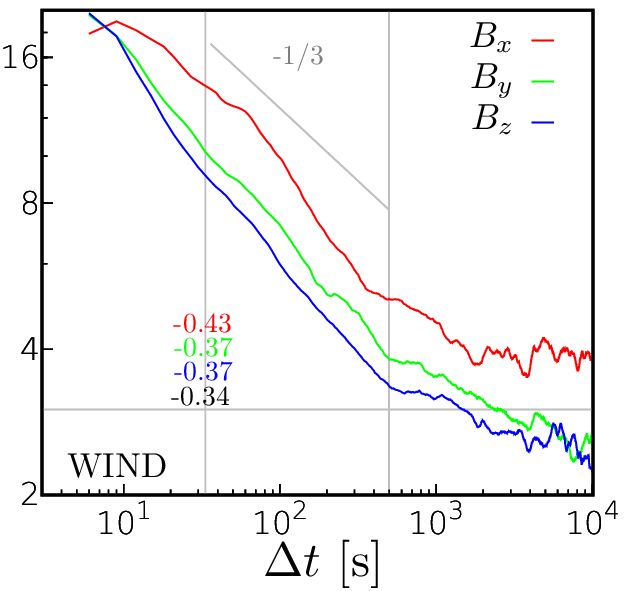}
\caption{Scale-dependent kurtosis of the magnetic field components, for the three samples. The fitted power-law exponents are shown for the MHD inertial regime (color coded), delimited by gray vertical lines. Power laws with exponent $-1/3$ are indicated as a reference (gray).}
\label{fig:kurtosis}
\end{figure*}
\begin{figure*}[h!]
\includegraphics[width=0.333\linewidth]{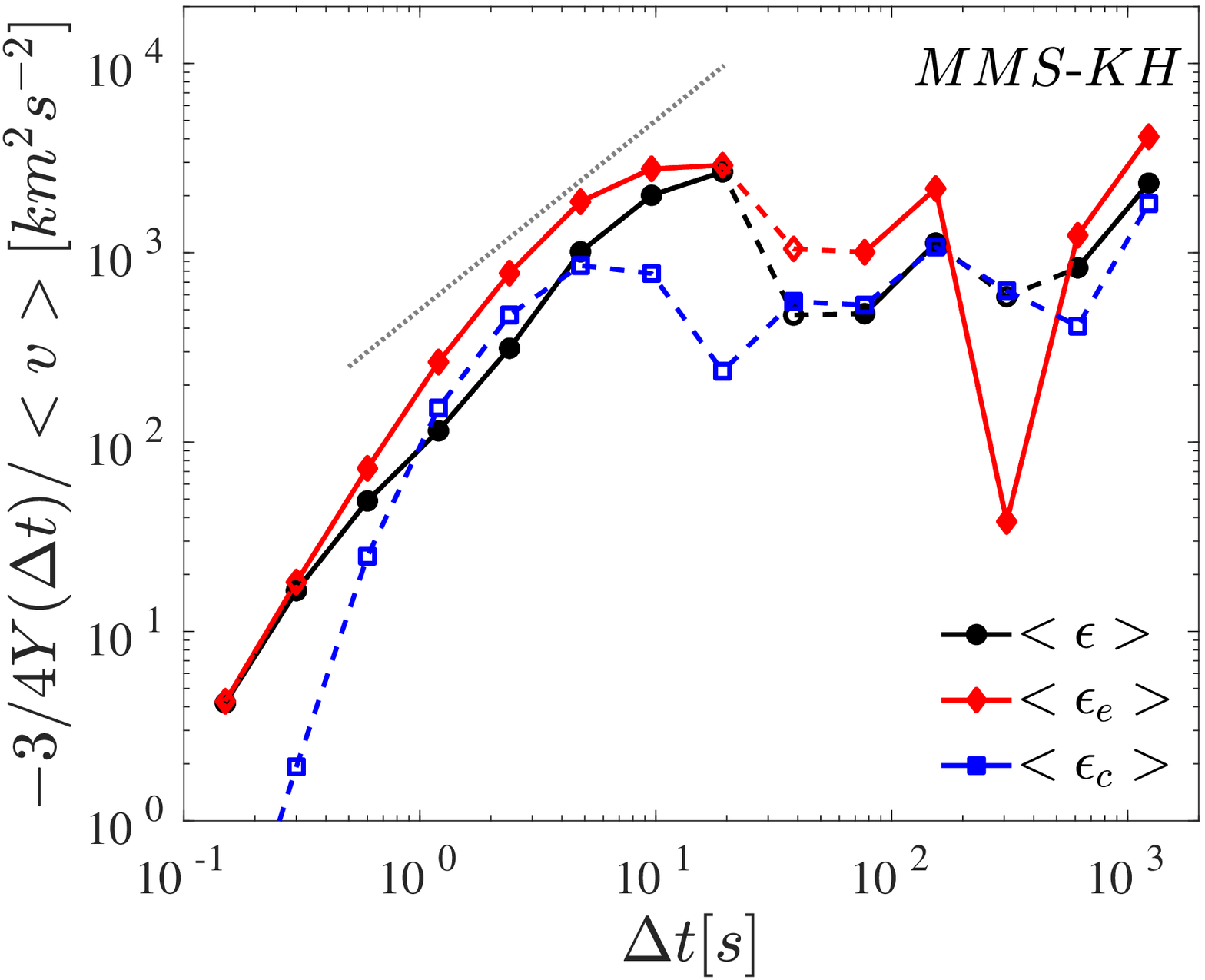}
\includegraphics[width=0.32\linewidth]{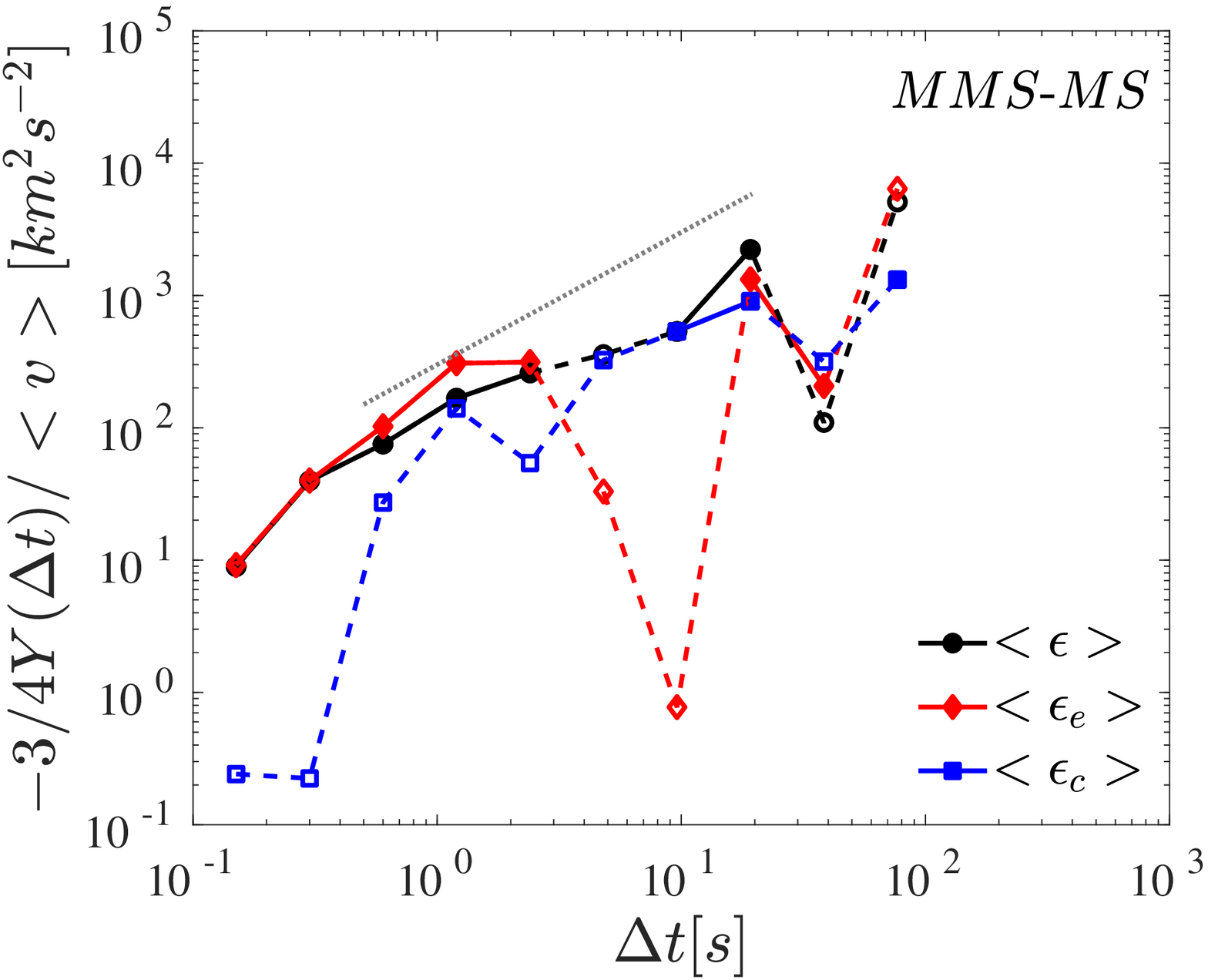}
\includegraphics[width=0.32\linewidth]{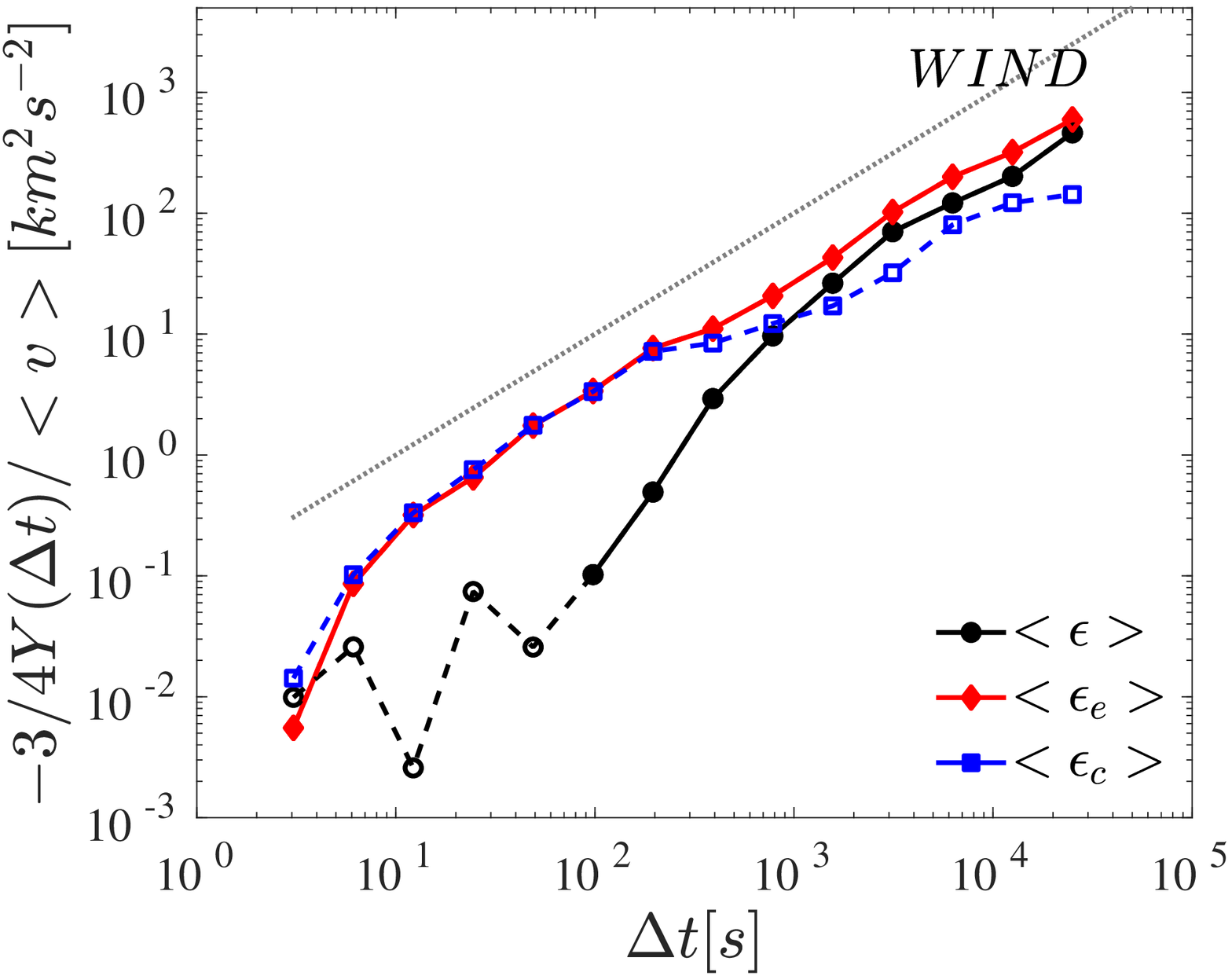}
\caption{The Politano-Pouquet law~(\ref{yaglom}) for the three samples, indicated as $\langle \varepsilon \rangle$ in the legend (black), and in terms of its averaged components $\langle \varepsilon_e \rangle$ (red) and $\langle \varepsilon_c \rangle$ (blue). Negative terms are represented as full lines, while the inverted positive terms as dashed lines. The thin grey lines represent linear scaling law, and are shown for reference.}
\label{fig:yaglom}
\end{figure*}

The formation of small-scale structures, typical of intermittency, is evidenced by 
the (roughly) power-law increase towards the small scales of the normalized fourth order moment (kurtosis) of the magnetic fluctuations, $K_i(\Delta t)=\langle \Delta B_i^4\rangle / \langle \Delta B_i^2\rangle^2$ (Figure~\ref{fig:kurtosis}), the subscript $i$ indicating the component $x$, $y$ or $z$. Note that the Gaussian value $K=3$ is observed for scales larger than the estimated inertial range (right grey vertical line). The Power-law decrease with the scale is a direct consequence of the structure function scaling in turbulent fields. The fitted scaling exponents are indicated in each panel, and are proportional to the degree of intermittency of the system~\cite{carbone}. For the solar wind data, where the turbulence is more developed, the power-law behaviour is more evident. Both the exponents and the small-scale magnitude of the kurtosis are in agreement with typical values for fast solar wind~\cite{livingreviews}. 
A shorter, less defined power-law scaling range, with slightly smaller scaling exponents, is observed in MMS-KH and MMS-MS data, suggesting a less developed intermittency in the younger turbulence of the shocked plasma. 
Similar results (not shown) were obtained through the standard analysis of the anomalous scaling of the structure functions~\cite{frisch}, fitted to a $p$-model~\cite{meneveau}, in the extended self-similarity approach~\cite{benzi}. 
After averaging over the three components (no major differences were observed), the magnetic field intermittency parameters $p$ are $0.67$, $0.79$, and $0.82$ for the MMS-KH, MMS-MS, and WIND samples respectively ($p$ lies in the interval $[0.5,1]$, with $p=0.5$ indicating absence of intermittency). These results show the strongly intermittent nature of the WIND sample,  the slightly less intermittent MMS-SH sample, and the weakly intermittent nature of the MMS-KH sample.

Finally, the Politano-Pouquet law~(\ref{yaglom}) can be estimated in the samples under study, both in terms of total energy transfer $\langle \varepsilon \rangle$, and in terms of the averaged components $\langle \varepsilon_e \rangle$ and $\langle \varepsilon_c \rangle$. The resulting scaling functions are shown in Figure~\ref{fig:yaglom} for the three intervals. None of the observed cases display a clear linear scaling. 
This might be due to the violation of the several requirements necessary for the Politano-Pouquet law to hold (e.g. incompressibility, isotropy, stationarity, large Reynolds number), to the presence of large-scale features advected by the flowing plasma, or simply to the lack of statistical convergence of the third-order moment, due to intrinsic finite-size limitation of space data. 
The challenging observation of the linear law in solar wind was already noticed using Ulysses data~\cite{prl,raf}.

In the MMS-KH data a power-law scaling slightly different from the expected linear relation is suggested. 
In WIND, there is evidence of linear scaling of the two components separately, while their combination does not display the predicted linear dependence. 
In the MMS-MS sample, the third-order moment and its components experience multiple sign inversions, possibly due to the finite size sample and to the expected poorly developed turbulence, evidenced by the steeper magnetic spectral exponent, as discussed above.

The intervals used in this work appear thus characterized, to different degrees, by the presence of an inertial range of turbulent, intermittent fluctuations, with a roughly defined energy cascade leading to the formation of small-scale structures.

\section{Results}
\label{results}

\begin{figure*}
\includegraphics[height=0.28\linewidth]{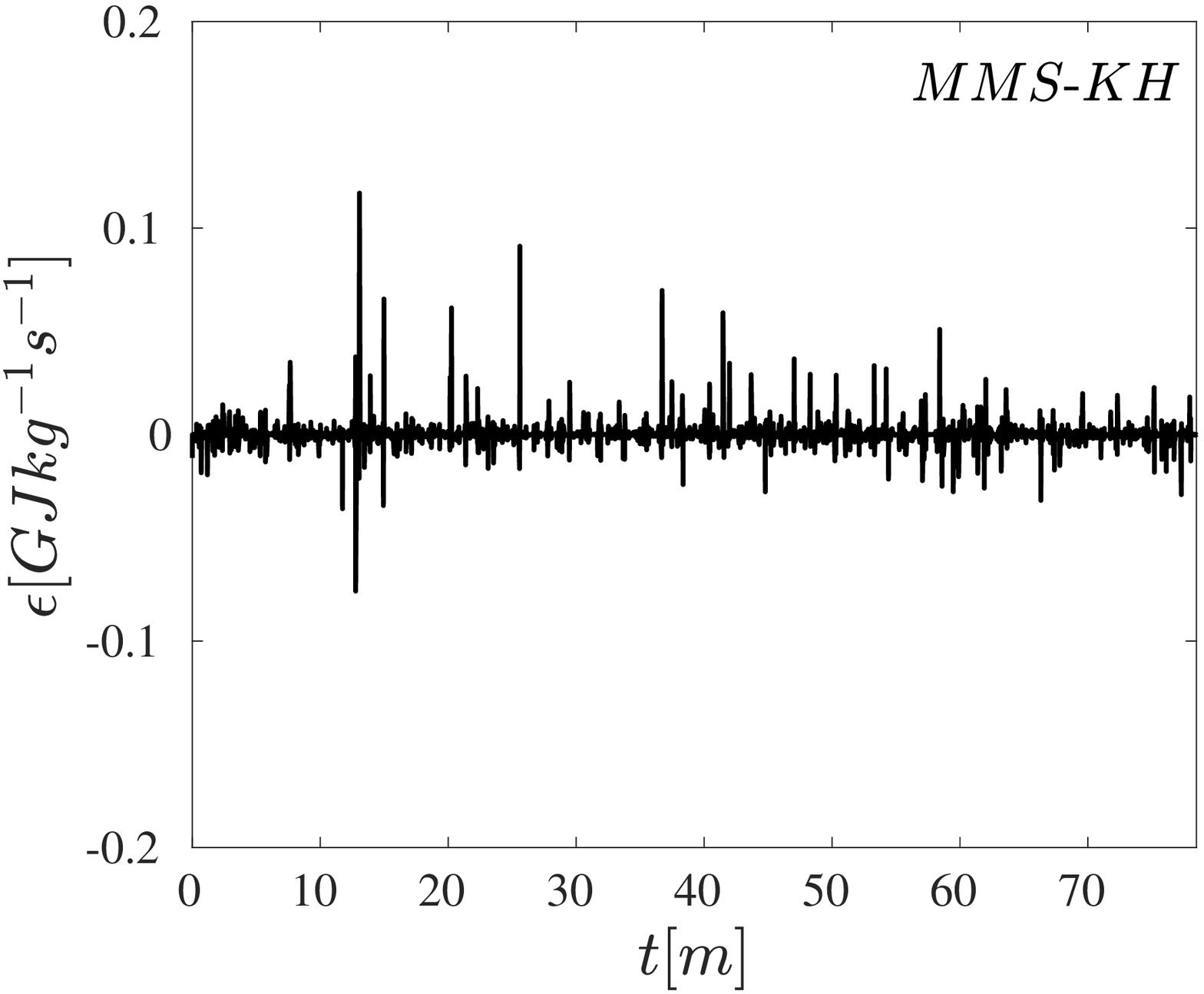}
\includegraphics[height=0.28\linewidth]{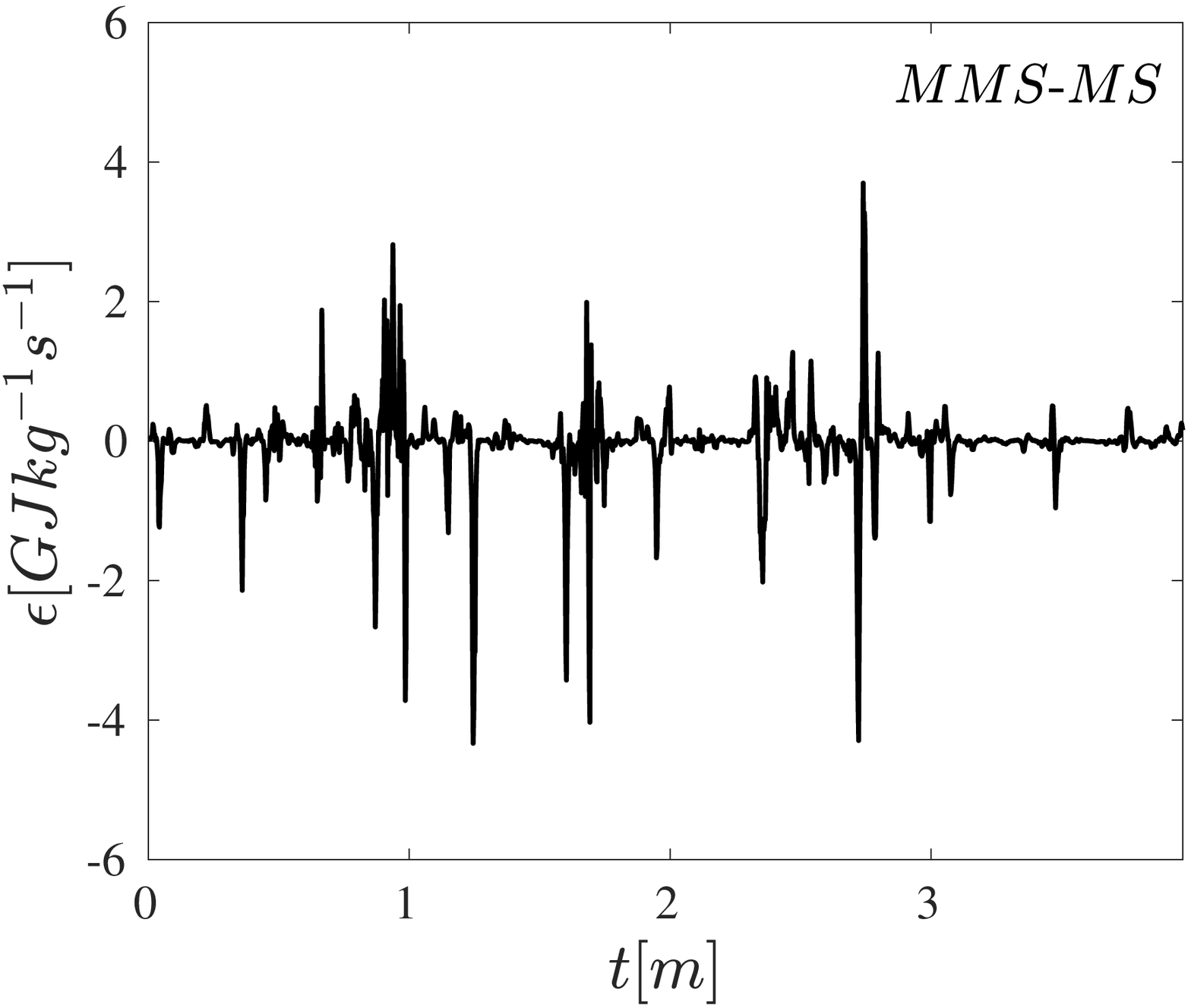}
\includegraphics[height=0.28\linewidth]{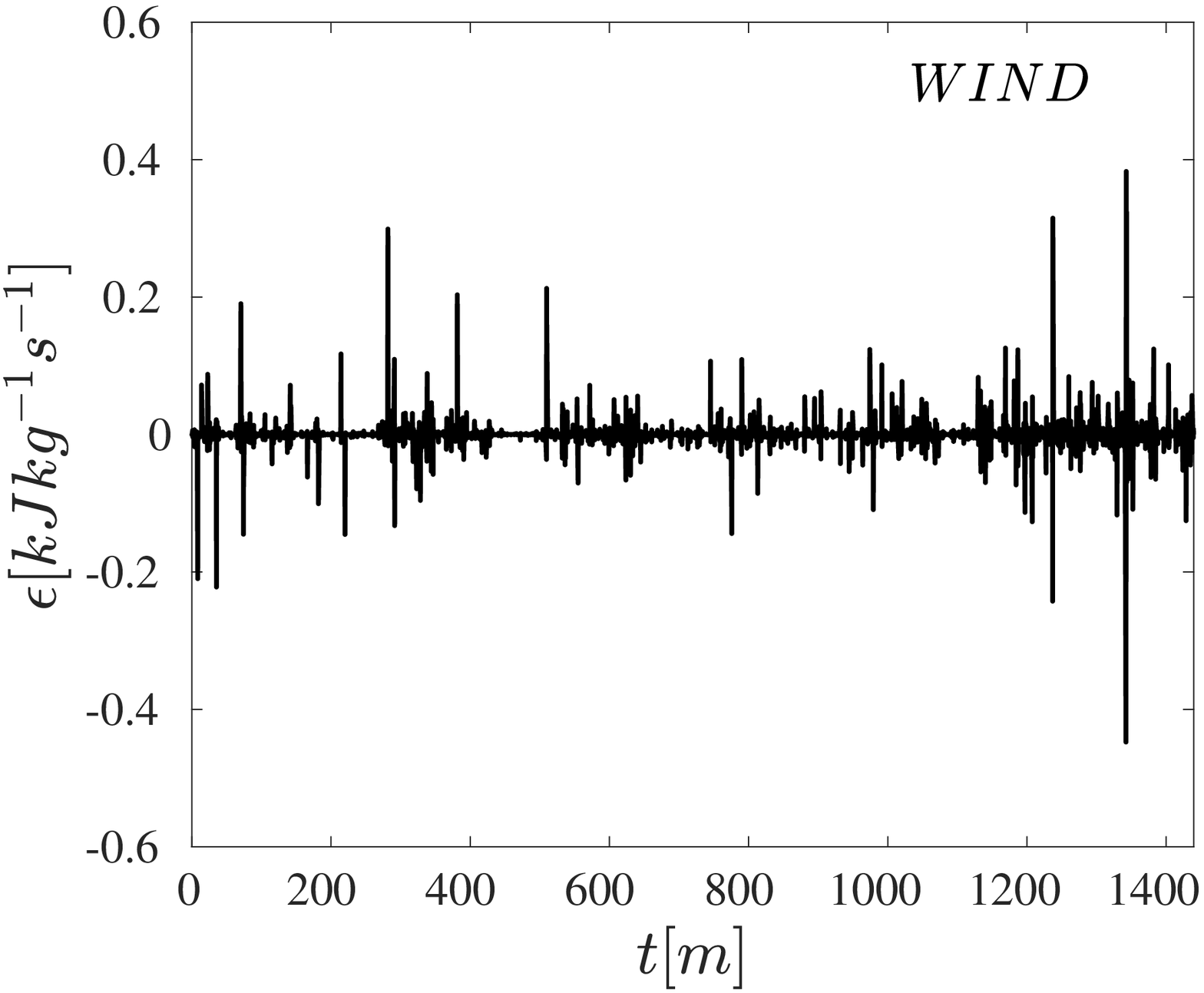}
\caption{The local proxy $\varepsilon$ as a function of time, estimated at the scale $\Delta t=1.2$s for the two MMS intervals, and at $\Delta t=6$s for WIND. Note that in the MMS-KH sample, the displayed signal results from the aggregation of the 53 separated sub-intervals, and has been displayed continuously in order to highlight its global properties.}
\label{fig:fluctuations}
\end{figure*}

From the MMS and WIND measurements described above, we have computed the LET, examples of which are shown in Figure~\ref{fig:fluctuations}.
In this work we will use the LET at scale of $1.2$ s, which is still inside and near the bottom of the MHD inertial range, where the Politano-Pouquet law is valid~\cite{sorriso2019}.
The proxy has the typical behavior of intermittent dissipation in turbulence~\cite{frisch}, with the presence of intense bursts of energy flux alternating with quieter regions.
The cancellation analysis described in Section~\ref{methods} was then performed on the signed fields $\varepsilon$, $\varepsilon_e$ and $\varepsilon_c$, as obtained from the different data sets considered for this study. A range of time-scale separations within the inertial range was considered, so that we have estimates of the LET for different scales $\Delta t$ within the turbulent cascade.  

\begin{figure*}
\begin{center}
\includegraphics[width=0.33\linewidth]{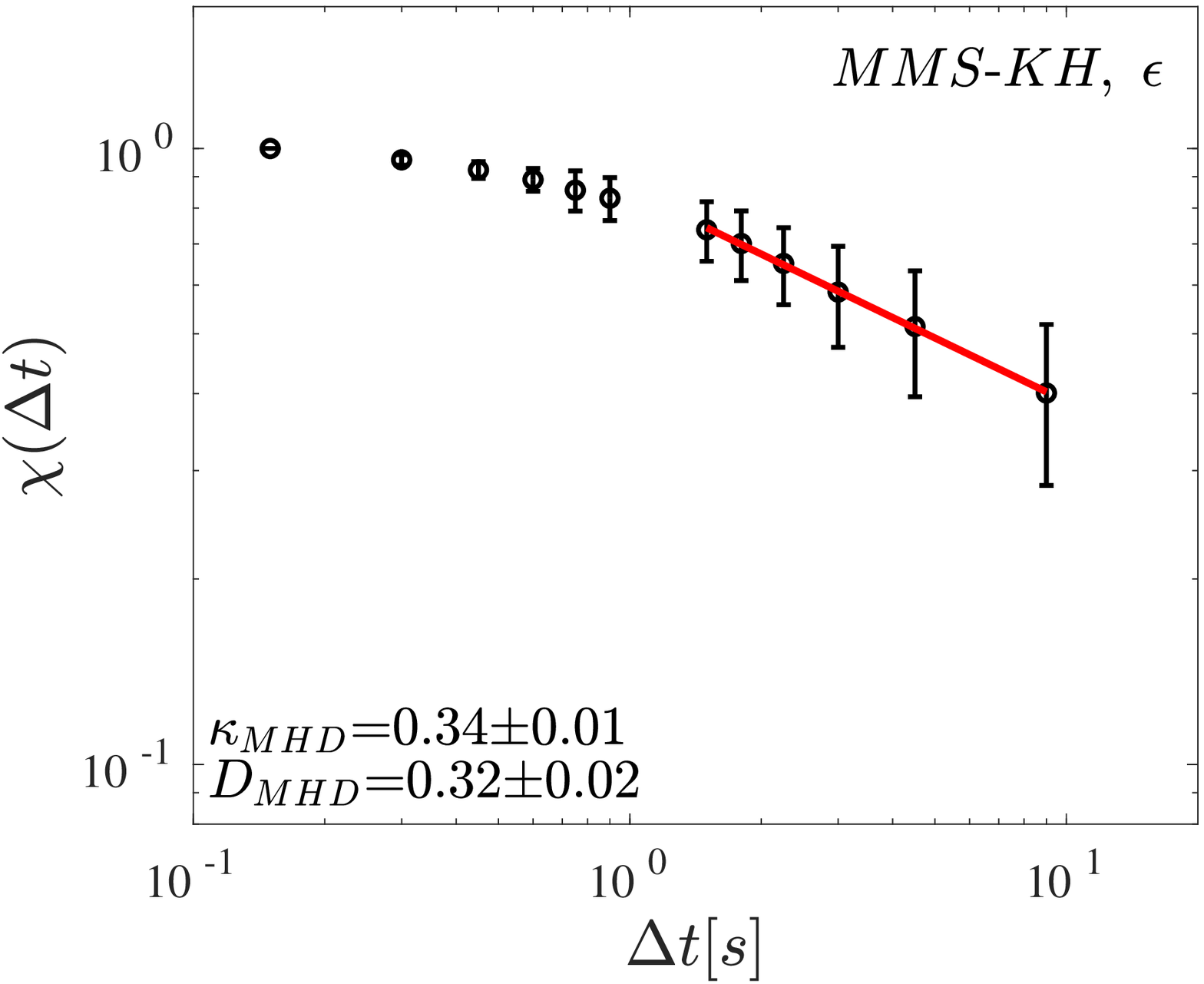}
\includegraphics[width=0.32\linewidth]{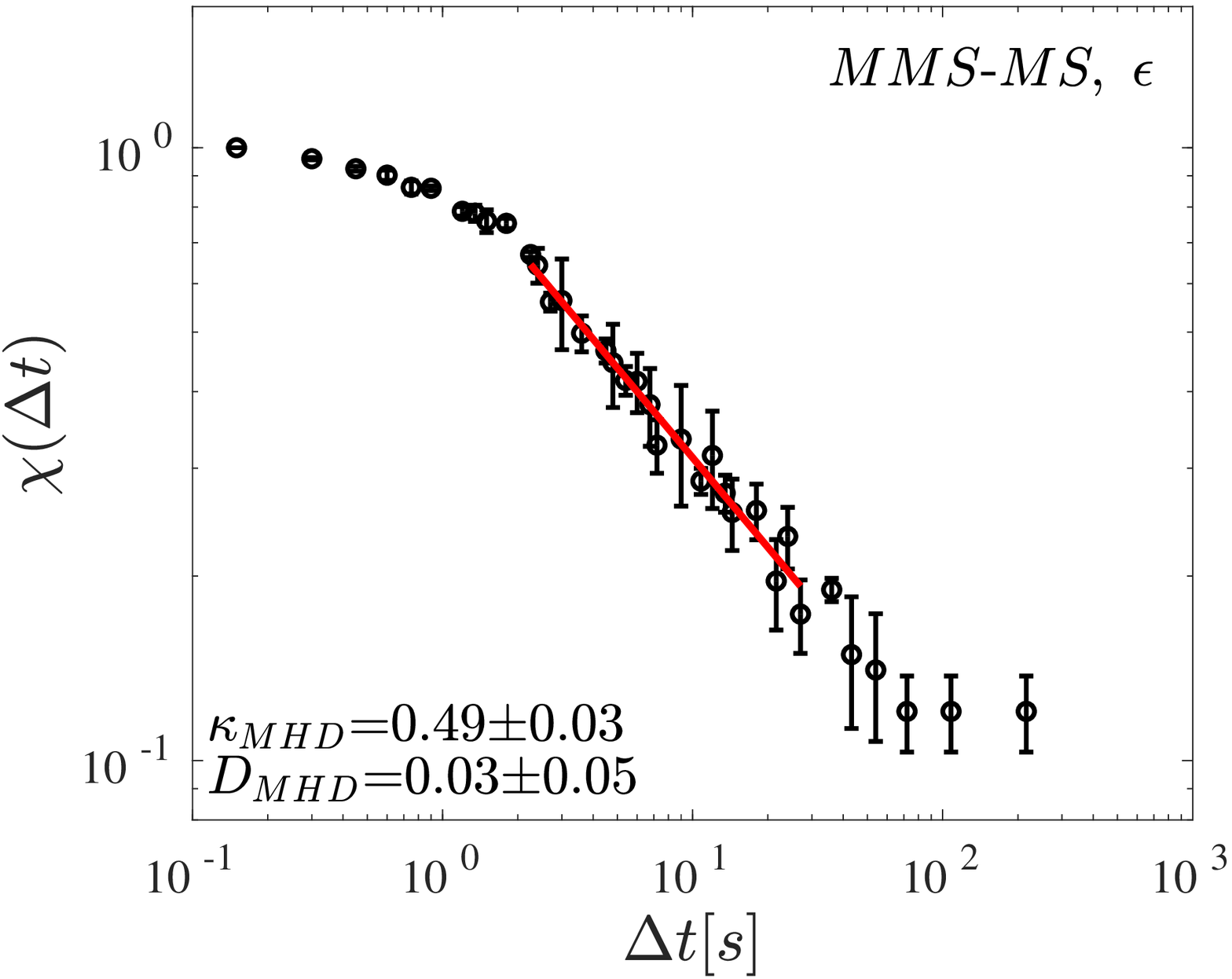}
\includegraphics[width=0.32\linewidth]{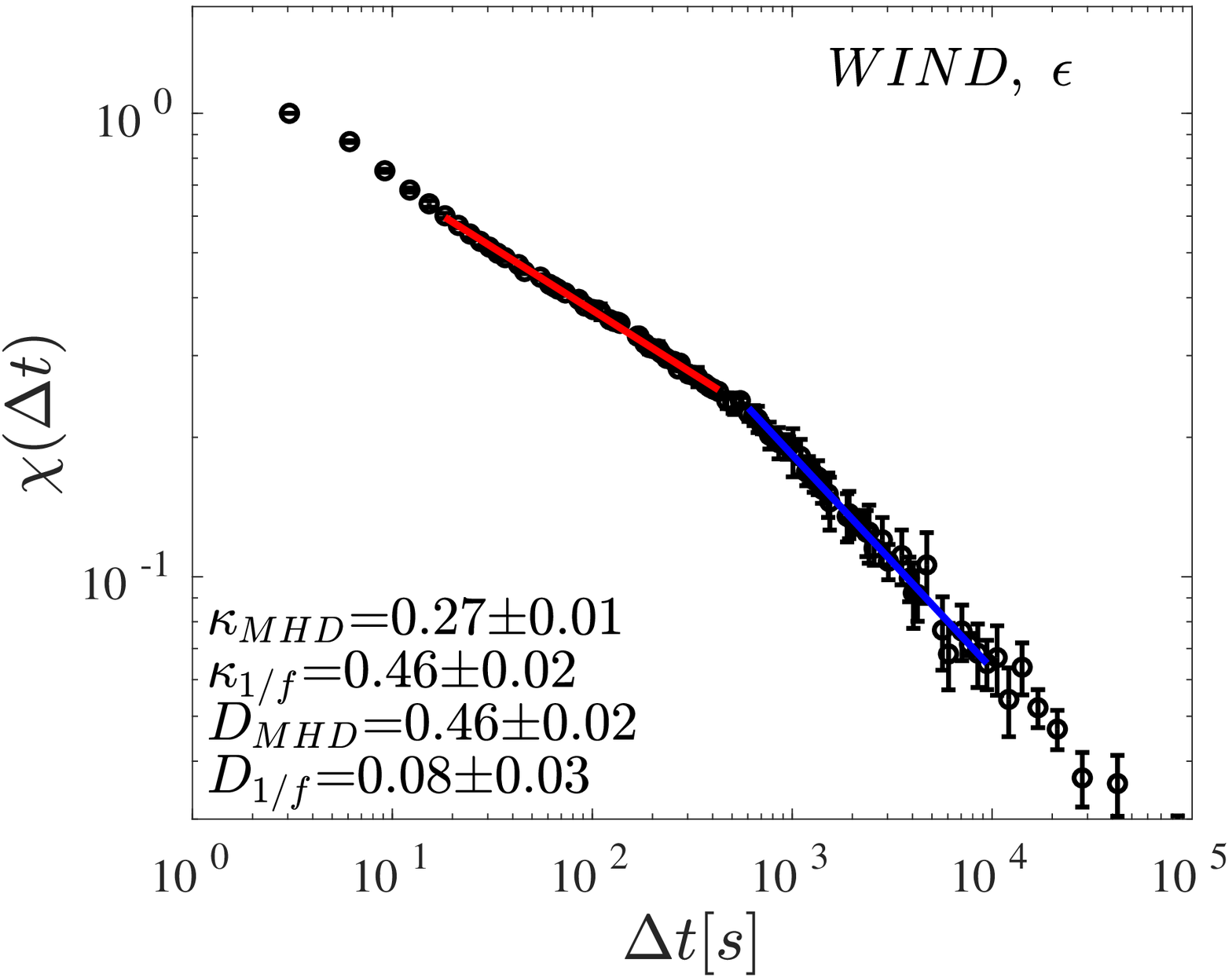}
\includegraphics[width=0.33\linewidth]{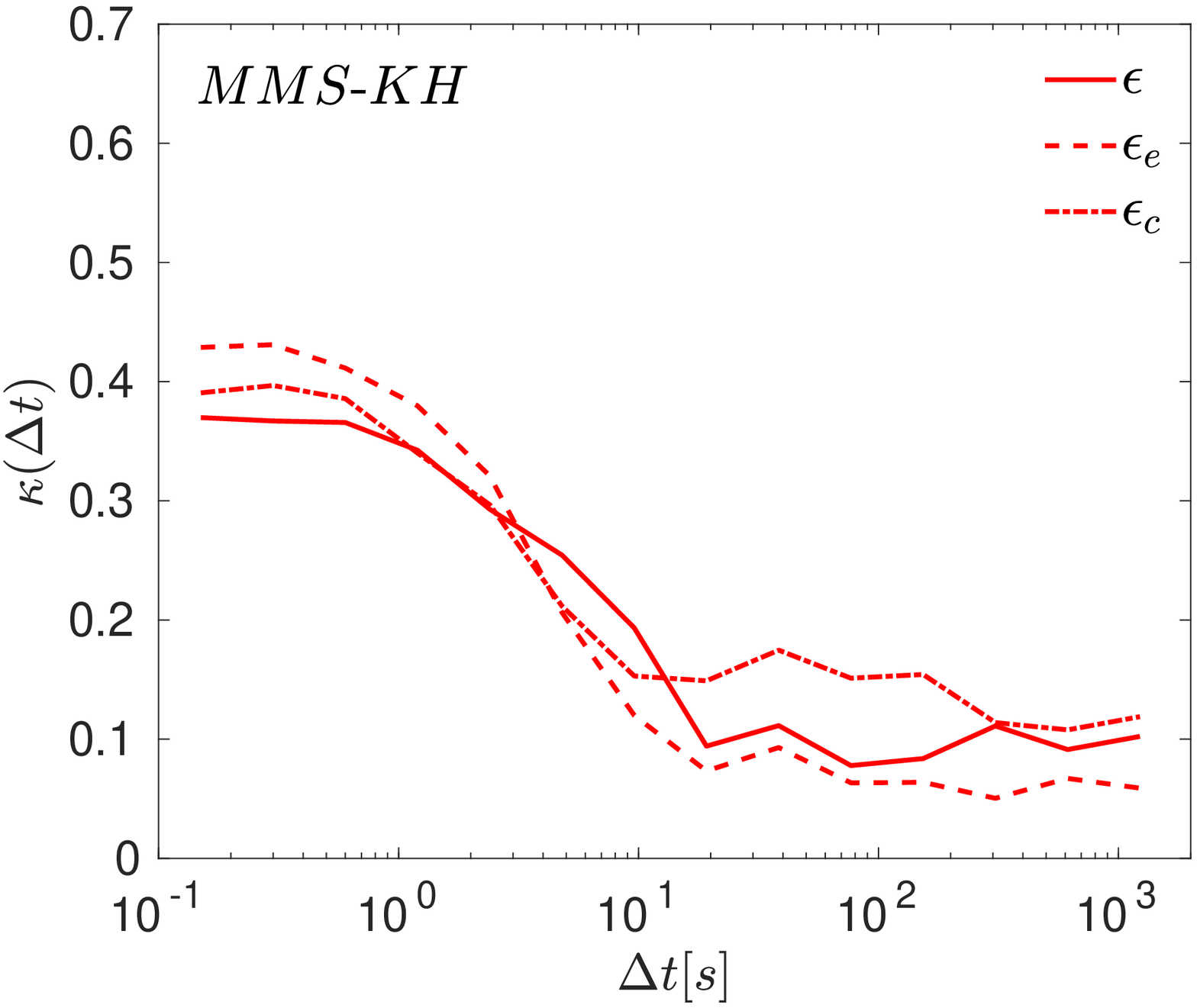}
\includegraphics[width=0.31\linewidth]{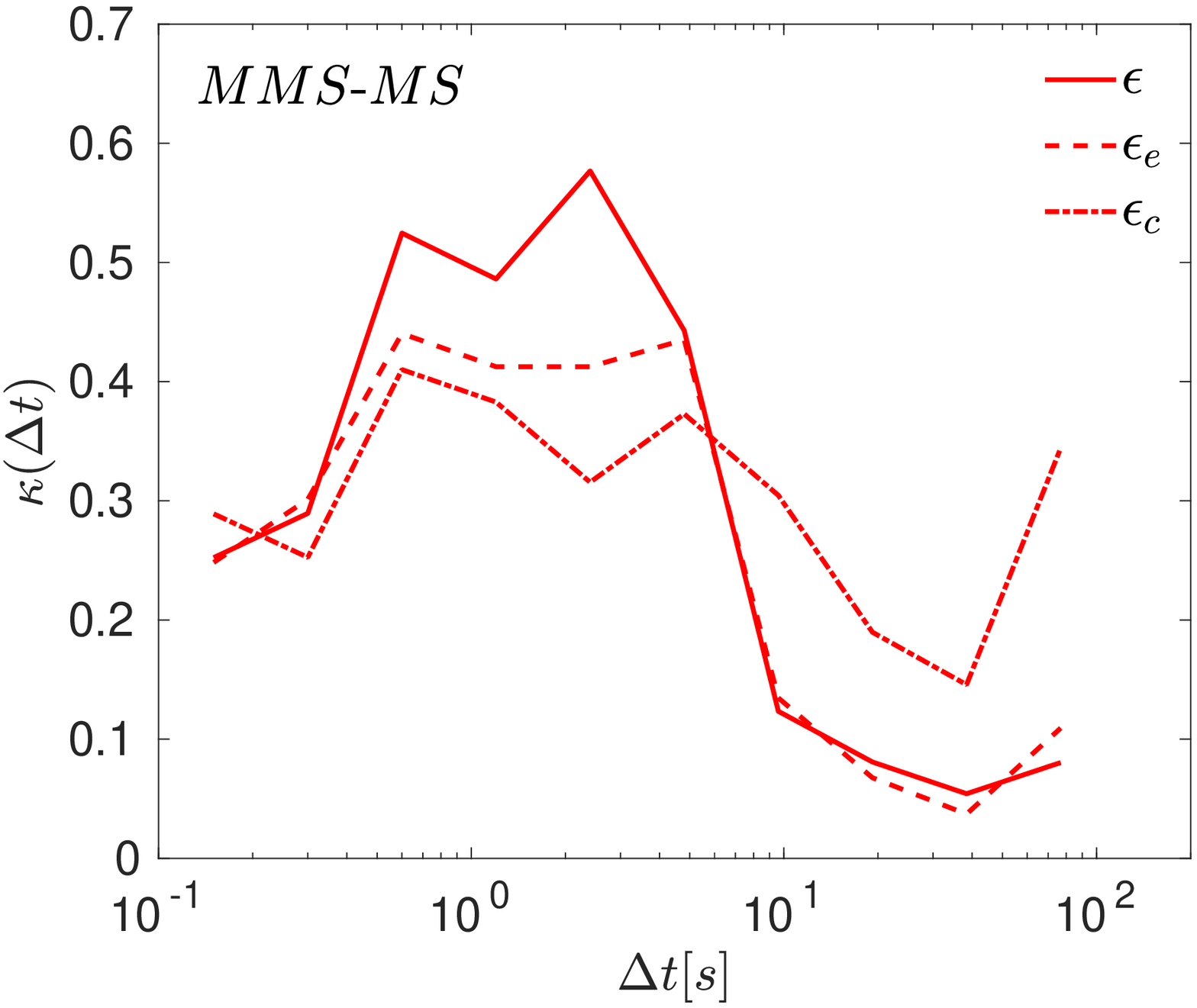}
\includegraphics[width=0.31\linewidth]{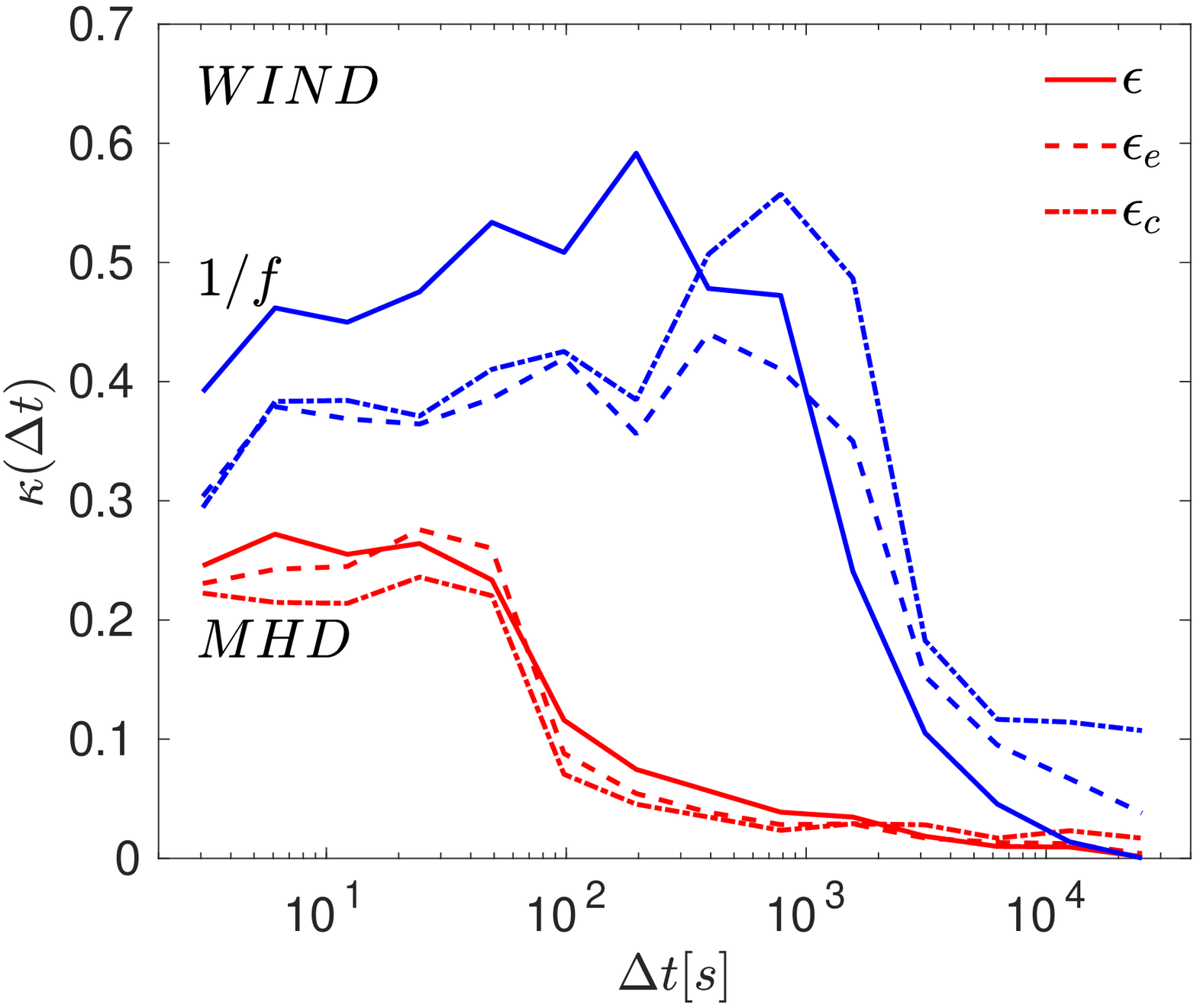}

\caption{Top row: the cancellation function $\chi(l)$ of the LET $\varepsilon$ for the three data sets. Power-law fits are indicated in the range of scale roughly corresponding to the spectral inertial range. Bottom row: the scale-dependent cancellation exponent $\kappa$, as computed from the fit in the MHD range of the cancellation function $\chi(l)$ of the three LET proxies $\varepsilon$ (full lines), $\varepsilon_e$ (dashed lines) and $\varepsilon_c$ (dash-doted lines), shown as red lines for the three data sets. In the right panel (WIND), the blue lines (same styles as above) refer to the proxies estimated in the $1/f$ range.}
\label{fig:chi}
\end{center}
\end{figure*}

Top panels of Figure~\ref{fig:chi} shows three examples of scaling of the cancellation function $\chi(l)$ for the LET proxy $\varepsilon$, computed using the field fluctuations at a scale $\Delta t$ near the bottom of the inertial range, as indicated in each figure. Each example refers to one of the three data sets studied in this paper. 
Power-law scaling can be easily identified in a region roughly corresponding to the respective inertial range of MHD turbulence (see Figure~\ref{fig:spectra} for comparison). 
In the WIND data, a possible secondary power-law scaling is observed in the large-scale range $\Delta t \gtrsim 10$ m, where spectra usually decay as $1/f$ (see in Figure~\ref{fig:spectra} the large-scale spectral break at $f\simeq 0.002$ Hz)~\cite{livingreviews}. 
On the other hand, the higher resolution of MMS data allows to highlight the presence of scaling in the ion range of scales (i.e. for $\Delta t \lesssim 1$ s, compatible with the spectral break visible near 1 Hz in Figure~\ref{fig:spectra}), where a different type of cascade may take place~\cite{alexandrova,stawarz,macek}. However, the scaling in this range should be studied in the framework of ion plasma physics, for example by including the Hall-MHD corrections to the Politano-Pouquet law~\cite{hellinger,ferrand}. This is left for future study. 
The cancellation functions have been fitted to power laws in the inertial range for all samples, and additionally in the $1/f$ scaling range for the solar wind data, providing the cancellation exponents $\kappa$, and thus the corresponding fractal dimensions $D=d-2\kappa$ (in this case $d=1$, so that the values of $D=1$ would indicate smooth, space-filling structures). Their values are indicated in the figure for some selected examples. 
A similar behaviour was robustly observed for all samples and all LET components, and at all scales $\Delta t$ within the inertial range, so that it is possible to compare the cancellation properties of the LET in the different cases.
\label{fig:kappa}
In the MMS-KH interval (left panel), the cancellation exponents near the end of the inertial range ($\Delta t = 1.2$ s) is $\kappa \sim 1/3$ for the three variables, a value indicating high complexity. This corresponds to fractal dimension of the order of $D\sim 0.33$, which is indicative of highly fragmented structures.
In the magnetosheath sample MMS-MS (central panel), at the same scale the exponent is closer to $\kappa \sim 1/2$, which is usually representative of random sign alternation, or absence of smooth, persistent structures of that scale. 
This is in agreement with the observed steep spectrum, indicative of the presence of weakly correlated discontinuities (or structures), and with the large kurtosis of this sample, which accounts for the broad presence of such structures.
%
In the fast solar wind sample measured by Wind, the scaling exponent of the total energy transfer proxy near the bottom of the inertial range ($\Delta t = 6$s) is $\kappa \sim 1/4$, which corresponds to the presence of structures of fractal dimension $D\sim 0.5$. This is in agreement with the typical observation of disrupted current sheets of solar wind intermittent turbulence~\cite{martin}, and confirms the fact that the turbulence is well developed in this fast wind stream, with strong intermittency. 
For the same interval, in the $1/f$ range of scales, a different fit of the cancellation function provides $\kappa \sim 1/2$, in excellent agreement with the uncorrelated nature of the Alfv\'enic fluctuations observed at such scales~\cite{livingreviews}.

Further information can be gained by observing the scale dependence of the cancellation exponent (or the corresponding fractal dimension). This can be obtained using the LET proxy estimated at different scales, using increments of the fields on variable scales $\Delta t$. Thus, for each scale, the LET and its components provide scale-dependent, local estimates of the turbulent energy flow.   
Results of cancellation analysis are collected in the bottom panels of Figure~\ref{fig:chi}, where the cancellation exponent $\kappa$ is shown for the three samples, and for the three variables. 
For all samples, at scales larger than the correlation scale (roughly $10$ s for both MMS samples~\cite{stawarz,sorriso2019} and about $30$ s for WIND, as evident from the spectrum and from the kurtosis) the cancellation exponent is $\kappa\lesssim 0.1$ (or $D\gtrsim 0.8$) for all fields, as expected for smooth, space-filling fluctuations.  
As the scale decreases, all samples display an increase of complexity, in a scale range roughly corresponding to the inertial range, where the intermittent structures are generated. Finally, a plateau or saturation seems to take place at or near the ion-scale spectral break. This could indicate that the intermittent structures have reached their stable geometry. 
However, this effect could also be due to the MHD nature of the LET proxies, which might be unable to properly capture further fragmentation of the fluctuations. The study of the ion range with the appropriate variable is left for future work. 

In the MMS-KH interval (left panel), the increase of complexity towards small scales is smooth and power-law like, and extends to the whole inertial range. The nonlinear cascade and the complex entanglement of positive and negative energy flux (proxies of the direct and inverse cascade, respectively) is thus beautifully captured by the LET in this sample. 
In the inertial range, all three variables (different lines) have similar exponents. 
The cross-helicity component $\varepsilon_c$ seems to provide slightly larger exponents than the energy component $\varepsilon_e$, suggesting that the energy transfer associated with current and vorticity structures occurs in a slightly smaller fraction of the volume (smaller fractal dimension). 

As for the KH interval, the WIND cancellation exponents obtained in the MHD range present similar exponents for the three MHD variables throughout the whole range of scales, indicating that the alternation of positive and negative energy flow is similar for the three proxies. This could be an indication of well developed turbulence, where a sufficient equilibrium between the competing terms in the cascade has been reached. The growth of the complexity roughly follows a power-law scaling, which confirms the excellent scaling properties of this sample, and that the LET proxies capture the sign complexity of the energy cascade. 

From the bottom panels of Figure~\ref{fig:chi}, it is evident that the overall behavior described above is roughly  coherent for the WIND (in the MHD range) and MMS-KH samples. This suggest the universality of the turbulent cascade mechanism, at least with respect to the sign-singularity properties, or, equivalently, to the fractal properties of the structures responsible for the energy transfer. The similarity is also corresponding to the presence of well-developed Kolmogorov spectra for both samples (see Figure~\ref{fig:spectra}), and to the power-law scaling of the kurtosis describing intermittency. In particular, the small-scale limiting values of $\kappa$ are larger for the MMS-KH sample than for WIND, in agreement with the more developed intermittency highlighted by the kurtosis and structure function analysis.

In the magnetosheath data, the increase of complexity of the positive-negative alternation for the total and structure-related proxies is sharper, less defined, and is observed right at the beginning of the spectral power-law range. 
The overall behaviour is different for the three proxies, with the total energy reaching a value of the exponent corresponding to uncorrelated, random fluctuations, while both components reach slightly smaller $\kappa$. 
Some degree of correlation is thus present in these two proxies, indicating the presence of extremely fragmented current, vorticity, and Alfv\'enic structures, whose superposition results in uncorrelated energy flux. The magnetosheath sample is thus probably characterized by less developed turbulence, corresponding to the steeper spectrum, and by the presence of small-scale structures, in agreement with the large kurtosis.  

Finally, in the WIND $1/f$ range, there is a similar trend as in the MMS-MS data, with smooth fluctuations at large scale, but the increase to uncorrelated, random values occurs sharply at the top of the inertial range, so that the energy flow associated with large-scale fluctuations clearly does not contribute to the energy cascade, as expected.

\section*{Discussion}
\label{conclusions}

The nature of the turbulent energy cascade has been analyzed in three samples of space plasmas by means of cancellation analysis applied to heuristic proxies of the local energy transfer. 
The analysis provided information on the sign alternation of the local mixed third-order fluctuations, which may be related 
to the fractal properties of the associated energy transfer and thus to small-scale dissipative processes. 
In two samples, namely in the solar wind and in the KH instability at the magnetospheric boundary layer, the turbulent cascade is well described by the proxies, and cancellation analysis captures the increasing complexity of the alternating positive and negative fluctuations. In these two samples, the energy is transferred to small scales eventually generating disrupted current and vorticity structures, as well as Alfv\'enic structures. The fractal dimension of these structures, obtained from the cancellation exponents, is indicative of a strong concentration of energy within a small fraction of the volume, typical of intermittency. 
The cancellation analysis of the magnetosheath sample studied in this work, on the contrary, provides an overall lower complexity estimate, which suggests the presence of less evolved turbulence, and lack of formation of well-structured energy channels. This is in agreement with the steeper spectrum and the more irregular Politano-Pouquet scaling law.

These results help to characterize the fluctuations that carry energy to smaller scales and provide the input or trigger for the activation of kinetic, dissipative processes in the small-scale  range~\cite{chen2019,klein,sorriso2019}. Moreover, the estimated one-dimensional projected fractal dimension provides information on the topology of the different types of fluctuations, namely of the current sheets, vorticity structures and Alfv\'enic fluctuations, that play an important role in the dissipation of the turbulence.

\begin{acknowledgments}
We acknowledge useful discussions with Denis Kuzzay, Lorenzo Matteini and Olga Alexandrova. 
FF acknowledges HPC@POLITO for computational resources.
LS and CV were supported by EPN Internal Project PII-DFIS-2019-01.
FF was supported by grant FOIFLUT 37/17/F/AR-B.
EY was supported by the Swedish Contingency Agency, grant 2016-2102.
CC was supported by STFC Ernest Rutherford Fellowship ST/N003748/2.
AG, KC, OK, DK and LS have received support from the Shota Rustaveli National Science Foundation Project No FR17$\_$279.
SP has received funding from the European Unions Horizon 2020 Research and Innovation programme under grant agreement No 776262 (AIDA, www.aida-space.eu).
The French LPP involvement for the SCM instrument on MMS was supported by CNRS and CNES.
MMS data are available at the MMS Science Data Center (https://lasp.colorado.edu/mms/sdc).
Wind data are available at CDAWeb (https://cdaweb.gsfc.nasa.gov).
\end{acknowledgments}


\end{document}